# A CENTURY OF BOSE-EINSTEIN CONDENSATION

Nick P. Proukakis
(Newcastle University, U.K.)

***Bose-Einstein Condensation is a phenomenon at the heart of many of the past century's most intriguing and fundamental manifestations, such as superfluidity and superconductivity: it was discovered theoretically some 100 years ago, and unequivocally experimentally demonstrated in the context of weakly-interacting gases 30 years ago. Since then, it has spawned a revolution in our understanding of fundamental phases of matter and collective quantum dynamics extending across all physical scales and energies, with unforeseen implications and the potential for envisaged quantum technological applications.***

-----------------------------------------------------------------------------------------------------------

100 years ago, in the early days of quantum mechanics, the first description of an intriguing, yet fundamental, physical phenomenon was formulated: **Bose-Einstein Condensation (BEC)** has since shaped physics research over many distinct subfields. Predicted theoretically during 1924-25, the – at the time unknown – Indian physicist Satyendra Nath Bose presented an alternative derivation of the black-body radiation spectrum for photons, using a novel way of counting states. Albert Einstein promptly extended this to non-interacting atoms in a series of papers (the latter published in early 1925). The main breakthrough of such an approach was its deviation from the established Boltzmann distribution of particles, predicting that non-interacting particles with integer spin preferentially occupy the same quantum state at appropriately low temperatures. This macroscopic occupation, now fittingly known as a Bose-Einstein condensate, leads to the emergence of collective dynamical behaviour and quantum coherence for any system under appropriate conditions. Like with most scientific discoveries, the path to its recognition has been somewhat 'bumpy' and not devoid of scientific viewpoint controversies.

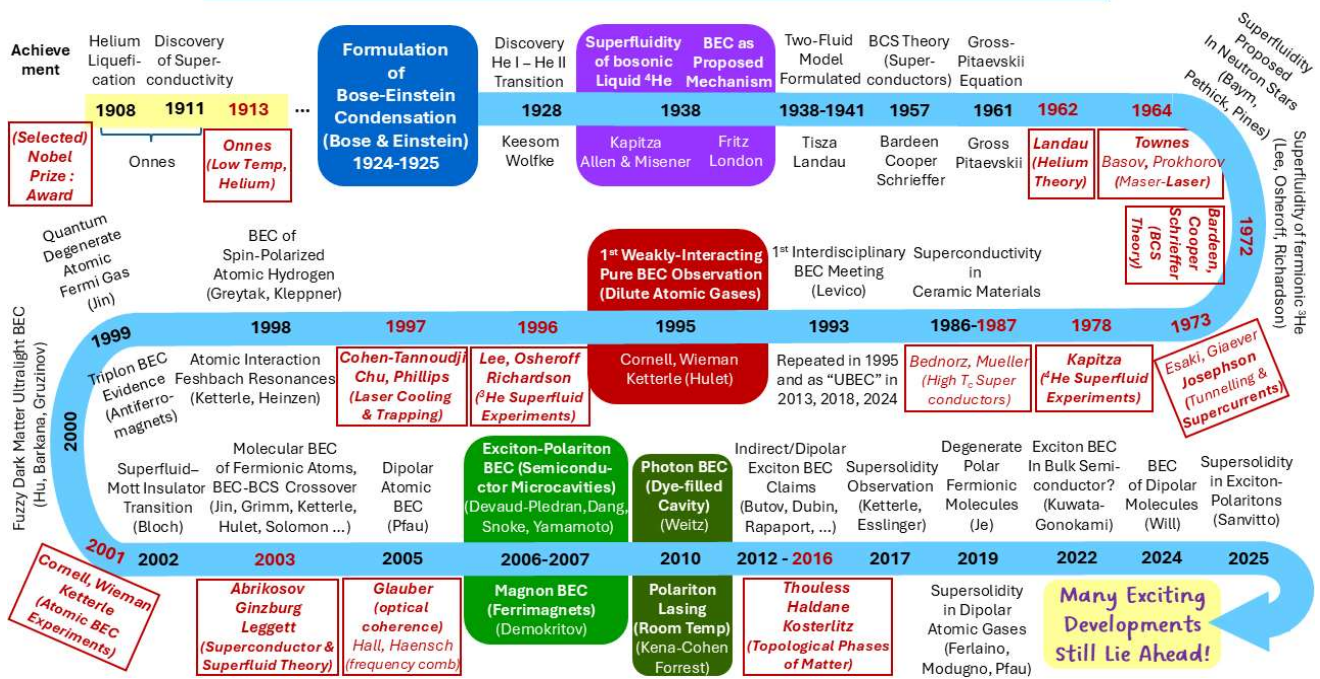

*Figure 1: Indicative non-exhaustive timeline of 100 Years of BEC. The graphic showcases a subjective selection of some of the main achievements, and corresponding Nobel Prizes (boxed text), related to BEC and macroscopic quantum coherence manifestations across different physical systems. Some emphasis is placed here on experimental advances facilitated by ultracold atoms, as the first controllable pure BEC system extensively investigated. Timeline indicates first credible claim in a given direction, even if not universally accepted; names are given as a non-exhaustive guide. To avoid diluting main BEC features, the present timeline does not include all relevant 'general' quantum advances (e.g. de Broglie's wavelength related to concept of quantum degeneracy, Pauli's exclusion principle relevant for fermionic pairing), broader inter-related physical advances (e.g. Meissner effect and London's phenomenological model, collective effects in nuclei, advances in neutron scattering critical to probing superfluid Helium, progress in various types of excitonic and photonic systems), or all theoretically-proposed types of condensation. For a somewhat up-to-date account of main physical systems exhibiting quantum-coherent BEC-based phenomena, see [1].*

This year we also celebrate 30 years since the first clear, unequivocal, demonstration of a pure Bose-Einstein condensate [2,3], the emergence of which I had the fortune to witness as a young PhD student in Oxford. Before commenting further on such experiments performed in the context of dilute, weakly interacting, trapped ultracold atomic gases, it is fitting to consider the emergence of BEC within the state of the art of 100 years ago.

BEC emerged as a model alongside two of the fundamental milestones of quantum mechanics: Louis de Broglie's matter wave hypothesis (1924), and Erwin Schrödinger's equation (1926). To add further context, the Heisenberg Uncertainty Principle was only introduced in 1927, Dirac's axiomatic approach to quantum mechanics published in 1930, and the famous Schrödinger cat thought experiment devised in 1935. It is thus evident that Bose-Einstein condensation is deeply enshrined within the development of the field of quantum mechanics. In this Comment, I briefly outline a subset of notable achievements in the century-long scientific journey of BEC, without having the pretence to provide a complete historical overview (relevant review references are given for each discussed topic).

**The Early Years of BEC and Superfluidity**

Despite its significance today, Bose struggled to get his initial paper published, only succeeding with Einstein's mediation. Even after Einstein's generalisation, the findings were not widely scientifically adopted. For example, in 1927 Uhlenbeck argued against the emergence of condensation of the ideal 'Bose-Einstein' gas in his PhD thesis, but this objection was publicly withdrawn following a scientific debate during the van der Waals centenary conference in late 1937 [4]. The following months were pivotal [5]: in January 1938 Kapitza, and independently Allen and Misener, published seminal works on the superfluidity of liquid Helium as back-to-back Letters to the Editor in Nature [6,7]. Just few months later, Fritz London established the first direct link [8] between BEC and experimental observables: His insightful application of the BEC model of a non-interacting gas, which ignores the very rich phenomenology of such a liquid, was the first key step to the establishment of the experimental evidence for the existence of a BEC state.

Subsequent theoretical work by Tisza and, independently, by Landau, led to the emergence of the two-fluid model of liquid Helium, which we now understand as the co-existence of an inviscid zero-entropy superfluid and an 'ordinary' (classical) fluid. Interestingly, such developments were based on conceptually different assumptions [5,9]: while London's (and subsequently Tisza's) picture was built upon the idea of a 'macroscopic quantum state', Landau did not believe in the role of BEC in the emergence of superfluidity, on the grounds that an ideal BEC description was irreconcilable with any model of a 'quantum liquid'.

The identification of superfluidity as the consequence of BEC became more established by the mid-1960's, following foundational works by Bogoliubov, Beliaev, Onsager, Penrose, Yang and others on Bose-Einstein condensation of *interacting* particles and the introduction of the concept of Off-Diagonal Long-Range Order (ODLRO). Interestingly, related heated debates persisted for many decades in the scientific community regarding the relation between the BCS (Bardeen-Cooper-Schrieffer) theory for superconductors and the BEC theory: although the foundational links between superfluidity and superconductivity were put forward by London in the late 1930's, the original BCS theory published in 1957 claimed such phase transition (in superconductors) not to be analogous to BEC. Nowadays, following key contributions by Blatt, Gorkov, Eagles, Leggett, Nozieres, Schmitt-Rink and others, we understand that a BEC phase was effectively 'hiding' within the original BCS model [10]. That is *not* to say that the highly distinct superfluid and superconducting systems exhibit the same macroscopic features, but rather that their macroscopic physics is largely shaped by a common *underlying* physical mechanism [11]. The complexity of the available BEC-related systems at the time (bosonic $^4$He superfluids, superconductors) did not facilitate an easy resolution to such theoretical puzzles, potentially even further fuelling the controversies. This led to a many-decade search for cleaner/purer systems exhibiting BEC, intended to overcome the difficulty associated with the depletion of the condensate due to strong interactions, which is significant for liquid Helium.

**Plausible BEC Candidate Systems: Particles, Composite Particles, or Quasiparticles?**

BEC requires an ensemble of bosonic particles whose de Broglie wavelength is comparable to the mean distance between them, allowing quantum statistical effects to emerge. As a result, practically any physical system can exhibit BEC under the appropriate conditions, leading to a diverse range of BEC systems across different physical scales, densities, and temperatures [10,12]. Details of individual systems exhibiting BEC, including those mentioned below, have been collected in a single volume [1] to which the reader is referred.

Already in his first 1925 paper, Einstein mentioned hydrogen, helium (and the electron gas) as the best possible candidates to observe his condensation phenomenon [4]. (Note that neither the Pauli exclusion principle, nor Fermi-Dirac statistics were known at the time). So, a natural first particle candidate to consider in the quest for a pure BEC was the bosonic Hydrogen atom, which remarkably maintains its gaseous phase down to absolute zero, when spin polarized. Despite coming very close to quantum degeneracy by the mid-1980's, and pioneering some techniques now used routinely in the ultracold atomic gas community, BEC in spin-polarised atomic Hydrogen was only achieved in 1998, three years after the ground-breaking pure BEC cold atom experiments.

Parallel developments explored whether weakly-interacting quasiparticles could also exhibit pure BEC. The first such suggestion, in 1962, was excitons (bound states of electron-hole pairs), with indirect excitons gradually emerging as a more promising platform due to their longer lifetimes. While such condensation proved elusive for a long time, significant experimental observations pointing to the emergence of BEC of indirect excitons have been reported in the past decade [13,14], including a recent report of direct observation of exciton BEC in a bulk semiconductor.

The question of weakly-interacting quasiparticle condensation triggered intense activity into other quasiparticle types, eventually leading to the unequivocal observation of exciton-polariton BEC in semiconductor microcavities in 2006: due to their significant advances since, such systems will be discussed below. Moreover, BEC of different types of 'magnon' and 'triplon' quasiparticle excitations was also observed in magnetic materials (see corresponding Chapters in [1] for details).

**The First Observation of a Pure BEC and the Versatility of Ultracold Atomic Gases**

The true significance of BEC became evident with its first unequivocal demonstration in a weakly-interacting dilute gas in 1995, which was achieved using optically and magnetically cooled and trapped alkali atoms at ultralow temperatures (typically below 1 μK). Following this experimental realisation, early characterizations [1-3,15] included a macroscopic demonstration of Heisenberg's uncertainty principle, studies of phase coherence, and collective motion, soon followed by concrete evidence of quantum fluid hallmarks like quantum vortices, persistent currents and Josephson superflow.

While I cannot do justice to such a rich field here, the maturity of ultracold atomic systems has enabled diverse research directions, including i) fundamental physics studies characterising coherence, transport, dissipation and phase transitions under various regimes, ii) nonlinear quantum dynamics and topological defects, and iii) novel physical regimes ideal for quantum applications and simulations. These systems offer unprecedented controllability in the lab, allowing real-time generation of different interaction strengths (attractive, or repulsive) and types (local, or long-range dipolar), and control over geometry (continuous, or discrete, also mimicking solid-state structure) and dimensionality.

Other notable developments include the generation of 'artificial' gauge fields, mixtures of same, or different, atomic species spanning across both bosonic and fermionic superfluids, and the emergence of self-bound liquid-like droplets. Such control facilitates the theorist's dream of constructing and testing distinct Hamiltonians and physical scenarios, leading to controllable quantum simulators, or using such systems as analogue systems, e.g. for quantum gravity. Such control has also enabled the studies of two important crossovers: from gas-like to hydrodynamic behaviour, and from 'BEC-like' condensation of molecules formed by two atoms to a 'BCS-like' condensation of momentum-paired Cooper pairs of atoms [16]. As a result, the study of ultracold atoms has furthered our understanding of the relation between gaseous BEC and superfluid Helium, and between superfluidity and superconductivity.

Recent years have seen the realization of the elusive quantum regime of supersolidity, a spatially ordered system with superfluid properties. First discussed more than 50 years ago, initial claims of its observation in liquid Helium set-ups in the early 2000's were retracted. Such a counter-intuitive state has now been observed in a range of different configurations with ultracold atoms [17]. Furthermore, the field of ultracold quantum chemistry, especially with polar molecules, is rapidly expanding [18], with recent observations of BEC in dipolar molecules.

**Emergence of BEC in Photonic Systems:**

Another major landmark is the observation of BEC of quasiparticles in a controllable open quantum system, first realised in 2006. The system consists of exciton-polaritons, that is 'half-light-half-matter' quasiparticles arising from the strong coupling of excitons confined in quantum wells and photons confined in microcavities: Exciton-polaritons undergo a transition effectively corresponding to BEC at a much higher temperature than atomic gases, owing to the much lower quasiparticle mass; such a state has already been observed in many different materials [1,14,19]. Exciton-polariton BECs have facilitated complementary controllable studies to ultracold atoms, including for example analogue gravity experiments and an alternative platform for the observation of supersolidity.

Unlike ultracold atoms, exciton-polaritons require pumping to counteract quasiparticle decay, making them open quantum systems. However, the control of the quality of the sample, excitation scheme and cavity parameters offer some control between open and closed quantum system dynamics and the degree of thermalization. Moreover, such systems allow non-invasive real-time monitoring of the field, through the 'leaky' nature of the microcavity mirrors.

A somewhat related system involves the direct condensation of photons, with photon thermalisation facilitated by embedding the photons within a dye-filled cavity. Controlled photonic systems also enable studies of the crossover between BEC and lasing [1,14], linking superfluidity and superconductivity to laser phenomena. Finally, the somewhat analogous (semiclassical) condensation of classical light waves has also been observed.

The abundance of diverse systems of controllable weakly-interacting BECs of both particles and quasiparticles, combined with the thermodynamic equilibrium nature of liquid helium, and our rapidly-advancing theoretical understanding, continue to reveal subtle aspects of quantum fluids and gases, and the arising macroscopic quantum behaviour of many-particle systems, with potential for new discoveries.

**The Truly Ubiquitous Nature of BEC: BECs in the Cosmos?**

BEC phenomena are believed to occur in high-energy physics, astrophysical, and cosmological settings. For instance, superfluid behaviour of neutrons and protons in neutron stars may explain pulsar glitches through quantum vortices [20]. On a cosmological scale, BEC of ultralight bosons is considered a potential mechanism for understanding dark matter within the 'Fuzzy Dark Matter' scenario representing galaxies as pure BECs [21], with surrounding halos acting as incoherent particles.

All above manifestations – both observed and inferred ones -- highlight the ubiquitous nature of the BEC phenomenon. This was the focus of the 1$^{st}$ International Workshop on Bose-Einstein Condensation held in Levico Terme in 1993, with the extended family of proposed BEC systems documented in [22]. The 1995 follow-up meeting saw the first public announcement of the ground-breaking BEC observation by Eric Cornell and Carl Wieman. For the next decade cold-atomic systems, as the only controllable pure BEC experimentally available, dominated BEC research, at least until developments in controllable exciton-polariton and related systems. Nowadays, various conferences and workshops, whether system specific or more interdisciplinary, discuss BEC-related physics, reflecting the broader community's evolving focus. This shift is evident in the 2017 interdisciplinary BEC volume [1] and the themes of the 2024 Universal Themes in Bose-Einstein Condensation conference held few months ago at the ECT* in Trento [23].

**Striking Features of the BEC Phase: Quantized Vortices and Josephson Effects**

The remarkable features of quantum gases, fluids, and superconductors are related to the coherent macroscopic wavefunction with a unique phase, as established by the works of London, Onsager, Feynman, Anderson and others [11,24]. This phase is constant in a static system and varies uniformly in a moving superfluid. Observations of superfluidity include suppression of scattering below a critical speed and the appearance of quantum vortices. Unlike classical vortices, quantum vortices exhibit quantized flows around a density minimum. While random configurations of quantum vortices can emerge spontaneously as the system crosses the BEC phase transition [25] a striking manifestation is their arrangement in regular structures in the presence of rotation, in a manner mimicking solid body rotation. Interestingly, the 'workhorse' equation of weakly-interacting quantum gases, known as the Gross-Pitaevskii equation [15], was used to characterize quantized vortex lines in liquid Helium in 1961, following their experimental observation by Hall and Vinen in rotating superfluid helium

Following the observation of vortex lattices in liquid Helium, such quantum mini-tornadoes have been demonstrated in rotating ultracold bosonic atomic gases, and subsequently in fermionic atomic superfluids, providing additional unequivocal proof of their superfluid features. Vortices have also been observed in atomic supersolids and exciton-polariton BECs, while their predicted existence in neutron stars provides a mechanism for explaining pulsar glitches. In superconductors, vortices appear as magnetic flux tubes in Type II superconductors under a uniform external magnetic field.

Another consequence of the macroscopic wavefunction manifests itself in the Josephson effects, predicted within BCS theory to describe current flow across two superconductors separated by a barrier. While initially elusive for many decades in Helium, such tunnelling-induced flow of neutral current has been observed – beyond fermionic $^3$He and bosonic $^4$He – also in weakly-interacting bosonic and fermionic ultracold atomic gas superfluids, and exciton-polariton condensates, highlighting their underlying superfluidity.

**Outlook**

The control of macroscopic quantum flow facilitates applications: a well-known example is the Superconducting Quantum Interference Device (SQUID) first invented six decades ago to measure magnetic fields. Unsurprisingly, within the emerging realm of experimentally controlled weakly-interacting systems exhibiting macroscopic quantum coherence, such 'newer' and purer BEC systems are also naturally being actively driven towards quantum-technological applications, including precision measurements and quantum sensing, such as accelerometers, gravitometers, and rotation sensors. By analogy to the established fields of electronics and spintronics, this has led to the studies of closed circuits of cold-atom and exciton-polariton BECs being respectively described by some as 'atomtronics' [26] and 'polaritronics' [27].

Such developments further pave the way for precision measurements of fundamental physical constants, providing a new approach to address long-standing fundamental physics questions. Examples include BEC studies under microgravity conditions on the International Space Station [28] and the demonstration of a cold atom gyroscope in the China Space Station, with current plans for large-scale cold atom interferometers. The field has now matured to include commercial BEC products and remotely controlled BEC experiments for educational purposes [29].

A century of BEC has brought significant advances, ground-breaking experimental, methodological and theoretical tools and deeper understanding of fundamental physics, alongside a fantastic playground for scientific fun. While such advances have already been recognized by multiple Nobel Prizes (see Fig. 1), the rapid ongoing developments, and the promise of technological applications, cast little doubt that there are more such recognitions ahead.

In closing, let me offer a personal perspective, having joined this remarkable BEC journey and interdisciplinary community 3 decades ago as a young graduate student: 100 years of BEC offers a direct example of the transformative power of scientific curiosity and disciplinary research, promoting an obscure physical phenomenon to a multi-billion-pound industry. This should be enough to convince politicians, university managers, and the broader public of the pressing need for more support and recognition for open-ended 'blue sky' disciplinary research, that need not necessarily be measured against our 'traditional norms' of research impact. I would also like to offer this as encouragement to younger researchers, to continue pursuing 'esoteric' studies that could lead to the next major scientific discovery 10, 50, or 100 years from now.

**Acknowledgments:**

*I am grateful to Carlo Barenghi, Franco Dalfovo and David Snoke for feedback on an early version of this Comment. My presented viewpoint has been shaped through discussion with too many colleagues to mention here, but it is fitting to highlight the Lorentz Centre, The Pittsburgh Quantum Institute and the ECT\* for supporting interdisciplinary BEC workshops during 2013, 2018 and 2024. I also acknowledge financial support from the Alexander S. Onassis Public Benefit Foundation, the UK EPSRC, the Leverhulme Trust and the European Union's Horizon 2020 programme.*